\documentclass[10pt]{article}

\usepackage{graphicx,psfrag}
\usepackage{hyperref}
\usepackage{amssymb}
\usepackage{amsmath}
\usepackage{amscd}
\usepackage{amsthm}

\usepackage[left=2cm,top=2.5cm,right=2cm,bottom=2cm]{geometry}
\begin{document}

\title{Bianchi type II models in the presence of perfect fluid and \\anisotropic dark energy}

\author{Suresh Kumar \footnote{Department of Applied Mathematics, Delhi Technological University, Bawana Road, Delhi-110 042, India. E-Mail: sukuyd@gmail.com} \and \"{O}zg\"{u}r Akarsu \footnote{Department of Physics, Ko\c{c} University, 34450 {\.I}stanbul, Turkey. E-Mail: oakarsu@ku.edu.tr}}
\maketitle
\begin{center}
\vskip-1cm
\textit{}
\end{center}

\begin{abstract}
Spatially homogeneous but totally anisotropic and non-flat Bianchi type II cosmological model has been studied in general relativity in the presence of two minimally interacting fluids; a perfect fluid as the matter fluid and a hypothetical anisotropic fluid as the dark energy fluid. The Einstein's field equations have been solved by applying two kinematical ans\"{a}tze: we have assumed the variation law for the mean Hubble parameter that yields a constant value of deceleration parameter, and one of the components of the shear tensor has been considered proportional to the mean Hubble parameter. We have particularly dwelled on the accelerating models with non-divergent expansion anisotropy as the Universe evolves. Yielding anisotropic pressure, the fluid we consider in the context of dark energy, can produce results that can be produced in the presence of isotropic fluid in accordance with the $\mathrm{\Lambda}$CDM cosmology. However, the derived model gives additional opportunities by being able to allow kinematics that cannot be produced in the presence of fluids that yield only isotropic pressure. We have obtained well behaving cases where the anisotropy of the expansion and the anisotropy of the fluid converge to finite values (include zero) in the late Universe. We have also showed that although the metric we consider is totally anisotropic, the anisotropy of the dark energy is constrained to be axially symmetric, as long as the overall energy momentum tensor possesses zero shear stress. 

\begin{flushleft}
\end{flushleft}

\end{abstract}
\section{Introduction}
\label{intro}

The discovery of the current
acceleration of the Universe from the
Type Ia Supernova (SNIa) observations \cite{Riess98,Perlmutter99}
posed one of the most challenging problems in modern cosmology. This
acceleration of the Universe is, today, a well established fact that
is confirmed by various independent observational data, including
SNIa, CMB radiation, etc. However, this discovery can be maintained
in the context of general relativity (GR) by introducing a
mysterious kind of energy source called the dark energy (DE), that can
generate repulsive gravity (see references \cite{Copeland06,LiMiao} for
extensive reviews on DE). It is well known that perfect fluids with
a constant equation of state (EoS) parameter lower than
$-\frac{1}{3}$ are, today, usually considered in the context of DE in cosmology,
since they can give rise to an accelerating expansion. Beside such
perfect fluids with constant EoS parameter, various scalar field
models, which can conveniently be described by an isotropic and time
dependent EoS parameter, have been studied with an intention of
producing kinematics consistent with the observed acceleration of
the Universe \cite{Copeland06,LiMiao}. However, this is not the only
possible method of studying DE. It can sometimes be useful to use a
reverse method, i.e., one may seek for the solutions of the Einstein
field equations by introducing some kinematical ans\"{a}tze that are
consistent with the observed kinematics of the Universe, and
may then obtain and investigate the dynamics of the fluid as a
possible candidate of the DE. For instance, as an ansatz in the
context of latter method, Berman's law \cite{Berman83,Berman88} for
the Hubble parameter, that yields constant deceleration parameter (DP),
has been widely considered for obtaining accelerating cosmological
models explicitly in the framework of spatially homogenous spacetimes after the discovery of
the current acceleration of the Universe (e.g., see Refs. \cite{Kumar11a,Kumar11b,Kumar11c} for recent studies on constant DP).

The sources with anisotropic stresses have been rarely studied in cosmology until recent years. This might be because of the expectation that a dominance of an anisotropic stress would give rise to an anisotropic expansion contrary to observations which favor an Universe that expands isotropically. One may see references \cite{Barrow97,Barrow99} for the effects of some known anisotropic stresses (such as magnetic fields, collisionless relativistic particles, hydrodynamic shear viscosity etc.) on cosmological evolution. On the other hand, the recent developments in cosmology increased the interest in anisotropic stresses particularly in the context of DE. Studies focusing on DE, including those in the framework of spatially homogenous but anisotropic spacetimes, have generally
assumed that DE yields an isotropic pressure and obeys a simple EoS
 in the form $p=w\rho$, where $\rho$ is the energy density, $p$
is the isotropic pressure and $w$ is the EoS parameter, which is not
necessarily constant. However, various discussions
\cite{Bennett03,Oliveira,Schwarz04,Cruz,Hoftuft09,BennettHill} that
focused on whether the Wilkinson Microwave Anisotropy Probe (WMAP) data \cite{Komatsu10,Komatsu09,Spergel07,Peiris03} for more successful explanation need a Bianchi
type morphology rather than Robertson-Walker (RW) type,  have promoted the
general interest not only in the Bianchi type cosmological models
but also in the possibility of anisotropic nature of the DE. For
instance, recently, Campanelli et al. \cite{Campanelli11a}
investigated the possibility that a DE component with anisotropic
EoS could generate an ellipsoidal Universe with the
correct characteristics to explain the low quadrupole in the CMB
fluctuations and proposed that an ellipsoidal Universe allowed
better matches with the large-scale CMB anisotropy data.  This
conclusion is also consistent with their more recent study
\cite{Campanelli11b}, where they analyzed the magnitude-redshift
data of SNIa included in the Union catalogue by considering Bianchi type I spacetime in the presence of a DE fluid with anisotropic EoS. In this study they concluded that supernova data are
compatible with a standard isotropic Universe, however a large level
of anisotropy, both in the geometry of the Universe and in the
EoS of DE, is allowed. Most recently, while this paper is in progress, Axelsson et al. \cite{Axelsson} studied possible imperfect nature of DE and the large-angle anomalous features in the CMB by considering the 7-year WMAP temperature observations data. Their results also motivate the further investigation on the possible origin and constraints of imperfect source terms including imperfect DE fluid in cosmology. Following such studies, several cosmological models
\cite{Koivisto08a,Rodrigues,Koivisto08b,Koivisto08c,Akarsu10a,Akarsu10c}
that introduce anisotropic DE candidates to break the spatial
isotropy of the Universe, which is generally believed to be obtained
via inflation in the early Universe, have been proposed. On the
other hand, even if it turns out that, in the future, the RW spacetimes are exactly
consistent with the observational data, this does not necessarily
rule out the possible anisotropic nature of the DE. The reason being that
anisotropic energy sources do not necessarily promote anisotropy in
the expansion, and sometimes can even support isotropization \cite{Akarsu10b,Akarsu11}.

Maximal spatial symmetry of
the RW metrics does not allow the
theoretical study of cosmological models containing anisotropic
fluids as long as fluids are comoving. We indeed do not expect the fluids
to have bulk motion at cosmological scales, and therefore we must take
the spatially anisotropic spacetimes into account to be able to
study cosmological anisotropic energy-momentum sources such as
magnetic field, cosmic strings, as well as hypothetical fluids
considered in the context of DE. In the context of the latter method
we mentioned above, one may allow the fluid to be general as much as
the metric allows and then seek for the solutions by introducing
kinematical ans\"{a}tze for the evolution of the Universe rather than constraining the fluid from the
beginning, for instance, by assuming that the fluid is isotropic. To
obtain explicit cosmological models in GR, we do not
have a general metric for spatially homogenous but anisotropic
spacetime. Instead, we have nine of the Bianchi type metrics plus
Kanstowski-Sach metric that form the complete class of spatially
homogenous but not necessarily isotropic four-dimensional spacetimes
\cite{Ellis69,Ellis08}. Hence, choosing one of these ten metrics as
the geometry of the Universe gives opportunity to generalize the
perfect fluid representation of the energy-momentum tensor (EMT).
However, the symmetry properties of the chosen metric would still
constrain EMTs that are allowed. Therefore, it is important to study
cosmological models within the framework of different types of
spatially anisotropic spacetimes, for a better understanding of the
possibly anisotropic nature of the DE and its effects on the
evolution of the Universe. Exact general relativistic cosmological
models in the presence of anisotropic fluid, in the context of DE, using kinematical ans\"{a}tze  have been
studied in the framework of the following metrics:
Bianchi I \cite{Akarsu10a,Akarsu10c,SharifZubair10a,Kumar10a,Pradhan10x,Sharif11},
Bianchi $\rm{VI_{0}}$ \cite{SharifZubair10b,Adhav11a,Amirhashchi10,Katore11,Katore11b},
Bianchi III \cite{Akarsu10b,Yadav11,Pradhan11},
Bianchi V \cite{Yadav11x,Yadav11BV},
Kantowski-Sachs \cite{Adhav11b,Adhav11c,Katore11Kantowski2},
Bianchi VI \cite{Akarsu11}
and
locally rotationally symmetric (LRS) Bianchi type II \cite{Pradhan11BII}. Such models have also been studied in $f(R)$ theory of
gravity instead of GR within the framework of Bianchi $\rm{VI_{0}}$
\cite{SharifKausar11a} and Bianchi III \cite{SharifKausar11b}
metrics.

In this paper, we present general relativistic cosmological models within the framework of spatially homogeneous but totally anisotropic and non-flat Bianchi type II spacetime in the presence of two fluids, which are the minimally interacting perfect fluid that can be described by an isotropic and constant EoS parameter, and the dynamically anisotropic hypothetical fluid as the DE fluid. We make use of two kinematical ans\"{a}tze at the same time, to fully determine the Einstein's field equations. As the first kinematical ansatz, we consider the generalized Berman's law \cite{Berman83,Berman88} for the Hubble parameter, that yields constant mean DP and whose negative values generate accelerating volumetric expansion. A similar procedure was carried out by Singh and Kumar \cite{Singh06}, within the framework of LRS Bianchi II spacetime, in the presence of a single isotropic fluid with a constant EoS parameter. In a recent paper, Naidu et al. \cite{Naidu11} have studied an LRS Bianchi type II DE model in a scalar tensor theory with generalized Berman's law and variable EoS parameter of DE. We would also like to note that although the constant DP is not capable of representing the whole or a long period of history of the Universe at once, it can still be considered as an approximation for representing a particular period of the history of the Universe, for instance, the vicinity of the present time of the Universe. As the second kinematical ansatz, we consider the assumption first introduced by Bali and Jain \cite{Bali98}, that assumes one of the component of the shear tensor to be proportional to the mean Hubble parameter.

\section{The metric, field  equations, solution and some immediate results}
\label{sec:2}
We assume that the GR is valid at
cosmological scales and consider Einstein's field equations
\begin{equation}\label{eq1a}
 G^{j}_{i}\equiv R^{j}_{i} - \frac{1}{2} R \delta^{j}_{i}=  -\frac{8\pi G}{c^{4}}T^{j}_{i},
\end{equation}
where $G^{j}_{i}$ is the Einstein tensor, $G$ is the Newton's gravitational constant, $c$ is the speed of light in the vacuum,  $R^{j}_{i}$ is the Ricci tensor, $R=R^{i}_{i}$ is the Ricci scalar, $\delta^{j}_{i}$ is the unit four-tensor and $T^{j}_{i}$ is the overall EMT that describes the physical ingredients of the Universe.

We consider the spatially homogeneous but totally anisotropic and non-flat Bianchi type spacetime in the form
\begin{equation}
\label{eq1} ds^{2} = - c^{2} dt^{2} + A^{2} (dx - K zdy)^{2} + B^{2} dy^{2} + C^{2} dz^{2},
\end{equation}
where $A$, $B$ and $C$ are the scale factors and functions of $t$ only. $K$ is a constant with dimension $\textnormal{length}^{-1}$. Choosing $K=0$ one gets Bianchi type I metric. For studying within the framework of Bianchi type II metric, which we are interested here in this paper, it is usual to choose $K=1$. We also, for convenience, switch to natural units $c = 1$ and $8\pi G=1$.

We assume the overall EMT $T^{j}_{i}$ consists of two different components; $T^{({\rm{m}})\;j}_{\;\;i}$ that shall represent the matter fluids (dust, radiation etc.) including Cold Dark Matter (CDM) and $T^{(\rm{de})\;j}_{\;\;i}$ that shall represent the hypothetical fluid we consider in the context of DE, i.e.,
\begin{equation}
T^{j}_{i}=T^{({\rm{m}})\;j}_{\;\;i}+T^{({\rm{de}})\;j}_{\;\;i}.
\end{equation}
We do not expect the fluids to have bulk motion on cosmological scales, hence we assume that the fluids are comoving,
i.e., fluids' four-velocities are $u^{i}=\delta^{i}_{1}$ with $ u^{i} u_{i} =-1$. Most of the cosmological matter fluids possess isotropic pressure and can be represented by a constant EoS parameter. Hence, the EMT that represents matter fluid can be written as
\begin{eqnarray}
\label{3}
T^{({\rm{m}})\;j}_{\;\;i}&=&\text{diag}\;[-\rho^{({\rm{m}})},\;p^{({\rm{m}})},p^{({\rm{m}})},p^{({\rm{m}})}]\nonumber\\
                   &=&\text{diag}\;[-1,\omega^{({\rm{m}})},\omega^{({\rm{m}})},\omega^{({\rm{m}})}]\rho^{({\rm{m}})},
\end{eqnarray}
where $\rho^{({\rm{m}})}$ is the energy density, $p^{({\rm{m}})}$ is the isotropic pressure and $w^{({\rm{m}})}=\rm{constant}\geq 0$ is the EoS parameter of the matter fluid. However, the nature of the DE is still mysterious and we have neither \textit{a priori} nor observational reasons for confining ourselves to assume DE yields isotropic pressure. Because the fluids are co-moving, one may get DE that yields an anisotropic pressure by assuming that the pressure of the DE is proportional to its energy density and subsequently the EoS of the DE fluid is direction-dependent. Hence we represent the EMT of the DE fluid as follows:
\begin{eqnarray}
\label{4}
T^{({\rm{de}})\;j}_{\;\;i}&=&\text{diag}\;[-\rho^{({\rm{de}})},p^{({\rm{de}})}_{x},p^{({\rm{de}})}_{y},p^{({\rm{de}})}_{z}]\nonumber\\
&=&\text{diag}\;[-1,w^{({\rm{de}})}_{x},w^{({\rm{de}})}_{y},w^{({\rm{de}})}_{z}]\rho^{({\rm{de}})}\nonumber\\
&=&\text{diag}\;[-1,\omega^{({\rm{de}})}+\delta,\omega^{({\rm{de}})}+\gamma,\omega^{({\rm{de}})}]\rho^{({\rm{de}})},
\end{eqnarray}
where $\rho^{({\rm{de}})}$ is the energy density and $p^{({\rm{de}})}_{x}$, $p^{({\rm{de}})}_{y}$ and $p^{({\rm{de}})}_{z}$ are the directional pressures and $w^{({\rm{de}})}_{x}$, $w^{({\rm{de}})}_{y}$ and $w^{({\rm{de}})}_{z}$ are the directional EoS parameters of the DE fluid respectively along the $x$, $y$ and $z$ axes. In the third line we parametrize the second line by introducing the skewness parameters $\delta$, $\gamma$ that determine the deviations of the EoS parameters of the DE fluid on the $x$ and $y$ axes from the one on the $z$ axis. Further, because the observations do not rule out the possibility of a dynamical and anisotropic DE, we allow all the parameters relevant to DE fluid to be time-dependent.

In the presence of the EMTs given in (\ref{3}) and (\ref{4}), Einstein's field equations (\ref{eq1a}) corresponding to the metric (\ref{eq1}) lead to the following set of linearly independent differential equations:
\begin{equation}
\label{eq7}
\frac{\ddot{B}}{B} + \frac{\ddot{C}}{C} + \frac{\dot{B}\dot{C}}{BC} - \frac{3}{4}\frac{A^{2}} {B^{2}C^{2}} = -w^{({\rm{m}})}\rho^{({\rm{m}})}-(w^{({\rm{de}})}+\delta)\rho^{({\rm{de}})} \;,
\end{equation}

\begin{equation}
\label{eq8}
\frac{\ddot{C}}{C} + \frac{\ddot{A}}{A} +\frac{\dot{C}\dot{A}}{CA} + \frac{1}{4}\frac{A^{2}} {B^{2}C^{2}}=-w^{({\rm{m}})}\rho^{({\rm{m}})}-(w^{({\rm{de}})}+\gamma)\rho^{({\rm{de}})}\;,
\end{equation}

\begin{equation}
\label{eq9}
\frac{\ddot{A}}{A} + \frac{\ddot{B}}{B} +\frac{\dot{A}\dot{B}}{AB} + \frac{1}{4}\frac{A^{2}} {B^{2}C^{2}}=-w^{({\rm{m}})}\rho^{({\rm{m}})}-w^{({\rm{de}})}\rho^{({\rm{de}})}\;,
\end{equation}

\begin{equation}
\label{eq10}
\frac{\dot{A}\dot{B}}{AB} + \frac{\dot{B}\dot{C}}{BC} +\frac{\dot{C}\dot{A}}{CA} - \frac{1}{4}\frac{A^{2}}{B^{2}C^{2}} =\rho^{({\rm{m}})}+\rho^{({\rm{de}})}\;,
\end{equation}

\begin{equation}
\label{eq7a}
\frac{\ddot{B}}{B} - \frac{\ddot{A}}{A} +\frac{\dot{B}\dot{C}}{BC} - \frac{\dot{C}\dot{A}}{CA} -\frac{A^{2}}{B^{2}C^{2}} = 0,
\end{equation}
where an overdot denotes $d/dt$. We have a system of five linearly independent equations \eqref{eq7}-\eqref{eq7a} involving eight unknown variables,
namely, $A$, $B$, $C$, $\rho^{({\rm{m}})}$, $\rho^{({\rm{de}})}$, $w^{({\rm{de}})}$, $\delta$ and $\gamma$. Therefore, in order to
fully determine the system, we need three constraining equations. To do that, we shall make one assumption on the physical ingredients of the Universe and two assumptions on the kinematics of the Universe. Before doing so, we need to define some kinematical parameters of importance in cosmology.

The directional Hubble parameters, which express the expansion rates of the Universe respectively along the $x$, $y$ and $z$ axes, can be defined as follows:
\begin{equation}
H_{x}=\frac{\dot{A}}{A}\textnormal{,}\qquad H_{y}=\frac{\dot{B}}{B}\qquad \textnormal{and}\qquad H_{z}=\frac{\dot{C}}{C}.
\end{equation}
The mean Hubble parameter, which expresses the volumetric expansion rate of the Universe, can be given as
\begin{equation}
\label{HubbleDef}
H=\frac{1}{3}\frac{\dot{V}}{V}=\frac{1}{3}\left(H_{x}+H_{y}+H_{z}\right),
\end{equation}
where $V=ABC$ is the volume scale factor of the Universe. The dimensionless mean DP $q$, which tells whether the Universe exhibits accelerating volumetric expansion or not, is another important kinematical quantity and can be defined as
\begin{equation}
q=\frac{d}{dt}\left(\frac{1}{H}\right)-1.
\end{equation}
The Universe exhibits accelerating volumetric expansion if $-1\leq q<0$, decelerating volumetric expansion if $q>0$, and exhibits constant-rate volumetric expansion if $q=0$.

The anisotropy parameter of the expansion $\Delta$ is crucial when deciding whether the models approach isotropy or not and can be defined as
\begin{equation}
\Delta=\frac{2}{3}\frac{\sigma^{2}}{H^{2}}=\frac{1}{3}\sum_{i=1}^{3}\left(\frac{H_{i}-H}{H}\right)^{2},
\end{equation}
where ${\sigma}^{2}=\frac{1}{2}\sigma_{ij}\sigma^{ij}$ ($\sigma_{ij}$ is the shear tensor) is the shear scalar. $\Delta$ is the measure of the deviation from isotropic expansion, such that the Universe expands isotropically if $\Delta=0$. Considering the kinematics of the models, we say that the model eventually approaches isotropy continuously if i) $V\rightarrow\infty$ and ii) $\Delta\rightarrow 0$ as $t\rightarrow \infty$ \cite{CollinsHawking}. However, considering the physical ingredient of the models, we also expect the energy densities of the each fluids to be positive as $t\rightarrow \infty$ for a more realistic approach.

Our first constraint is on the physical ingredients of the Universe. In GR, twice-contracted Bianchi identity tells that the overall EMT is conserved, i.e., $T^{\;ij}_{\;\;\;;j} =0$;
\begin{equation}\label{eq7b}
\dot\rho^{({\rm{m}})}+3(1+w^{({\rm{m}})})\rho^{({\rm{m}})}H+\dot\rho^{({\rm{de}})}
+3(1+w^{({\rm{de}})})\rho^{({\rm{de}})}H+\rho^{({\rm{de}})}\left(
\delta H_{x}+\gamma H_{y}\right)=0.
\end{equation}
Regarding the constraint on the physical ingredients of the Universe, it is quite reasonable to assume that the fluids are minimally interacting, i.e., the EMTs of the two fluids are conserved separately, as done by the authors of Ref. \cite{Akarsu10a}. Hence, for the matter fluid we have
\begin{equation}
\label{2.12}
T^{({\rm{m}})\;ij}_{\;\;\;\;\;\;\;\;\;\;;j}=\dot\rho^{({\rm{m}})}+3(1+w^{({\rm{m}})})\rho^{({\rm{m}})}H=0,
\end{equation}
which also trivially gives
\begin{equation}\label{14a}
\rho^{({\rm{m}})}=c_{0}(ABC)^{-(1+w^{({\rm{m}})})},
\end{equation}
where $c_{0}$ is a positive constant of integration. Similarly, from the conservation law for the EMT of the DE fluid we have
\begin{equation}
\label{2.13}
T^{({\rm{de}})\;ij}_{\;\;\;\;\;\;\;\;\;\;;j}=\dot\rho^{({\rm{de}})}+3(1+w^{({\rm{de}})})\rho^{({\rm{de}})}H
+ \rho^{({\rm{de}})}\left(\delta H_{x}+\gamma H_{y}\right)=0,
\end{equation}
which cannot be solved for $\rho^{({\rm{de}})}$ trivially as can be done for the matter fluid above.

For the second constraint (the first kinematical constraint), we consider Bianchi spacetime generalization of Berman's special law \cite{Berman83,Berman88} of variation for the mean Hubble parameter that yields a constant value of  mean DP:
\begin{equation}
\label{eq14} H = \ell V^{-\frac{n}{3}},
\end{equation}
where $\ell>0$ and $n\geq0$ are constants. This law has been widely used for obtaining exact cosmological models in the context of anisotropic DE. This law gives us opportunity to study two kinds of volumetric expansion laws. From equations (\ref{HubbleDef}) and (\ref{eq14}), we obtain
\begin{equation}
\label{eq16}
V = (n \ell t + c_{1})^{\frac{3}{n}}\quad\textnormal{for}\quad n\neq 0,
\end{equation}
\begin{equation}
\label{eq17}
V= c_{2}^{3}e^{3 \ell t} \quad\textnormal{for}\quad n=0,
\end{equation}
where $c_{1}$ and $c_{2}$ are constants of integration. Thus, the law (\ref{eq14}) provides the
power-law (\ref{eq16}) and the exponential-law (\ref{eq17}) of volumetric expansions of the Universe.
One may check that $q=n-1$, hence the models for $0\leq n < 1$, i.e., $q<0$ can be considered in the context of DE. In this paper, we are particularly interested in the accelerating models in the context of DE, hence relatively later times of the Universe rather than the early Universe. Therefore, for convenience, we shall only discuss accelerating models and the relatively late time behavior of the Universe in detail. Recent observations suggest that today $q\simeq -0.7$ (e.g., see references \cite{Cunha,Li2011}), hence, assuming that the DP changes slowly, particularly the model for $n\simeq 0.3$ would represent the Universe in the vicinity of present time. The exponential model ($n=0$), on the other hand, would also represent the very late Universe.

As the third constraint (the second kinematical constraint) to close the system, we assume that the component $\sigma^{1}_{~1}$ of the shear tensor $\sigma^{j}_{~i}$ is proportional to the mean Hubble parameter $H$, following the study of Bali and Jain \cite{Bali98}.
This constraint leads to the following relation between the scale factors:
\begin{equation}
\label{eq12} A = (BC)^{m},
\end{equation}
where $m>0$ is a constant.

Introducing the three constraint equations, one can now obtain the exact solutions by starting with obtaining the scale factor $A$ on the $x$ axis explicitly by using (\ref{eq12}) and the definition of the mean Hubble parameter (\ref{HubbleDef}) with (\ref{eq16}) for the power-law solution and with (\ref{eq17}) for the exponential-law solution. Integrating equation (\ref{eq7a}) two times, we obtain the scale factor $B$ on the $y$ axis  in terms of $A$ and $V$ as
\begin{equation}
\label{eq19} B = c_{4} A \exp \left[\int \left\{\frac{1}{V}\int  \frac{A^{4}}{V} dt\right\} dt+c_{3} \int \frac{1}{V} dt \right],
\end{equation}
where  $c_{3} $ and  $c_{4} $ are constants of integration. Finally, one can obtain the scale factor $C$ on the $z$ axis by using $A$ and $B$ in (\ref{eq12}).

After obtaining all of the scale factors explicitly, one can obtain all other parameters by using the explicit functions of the scale factors and the energy density of the matter fluid (\ref{14a}) in equations (\ref{eq7})-(\ref{eq10}). However, before proceeding to obtain the complete exact solutions, here, we would like to discuss two immediate findings.

First, we would like to emphasize an important point that arises from the second kinematical ansatz, which gives (\ref{eq12}), and the condition for approaching isotropy continuously. Using the second kinematical ansatz
(\ref{eq12}) we obtain
\begin{equation}
\label{eqH12}
H_{x}=m\left(H_{y}+H_{z}\right).
\end{equation}
The condition for approaching isotropy continuously imply that all the directional Hubble parameters must evolve to an identical value as $t\rightarrow \infty$, i.e., $H_{x}=H_{y}=H_{z}$ at $t=\infty$. One may check from (\ref{eqH12}) that $H_{x}=H_{y}=H_{z}$ can be possible only if $m=\frac{1}{2}$. Hence, if $m\neq \frac{1}{2}$ one cannot obtain a model that can approach isotropy continuously, but a model that approaches isotropy only for a certain period of time and a model that recedes from isotropy continuously. However, we also note that $m=\frac{1}{2}$ is the necessary but still insufficient condition for approaching isotropy continuously. In other words, choosing $m=\frac{1}{2}$ does still not assure approaching to isotropy continuously but provides the possibility, i.e., if $m=\frac{1}{2}$, one still needs to make further analyses of the anisotropy parameter of the expansion whether it can satisfy or cannot satisfy the conditions for approaching isotropy continuously.

Second, we would like to notice a relation between the BII metric and the anisotropy of the DE fluid. One can observe that, from (\ref{eq7}), (\ref{eq8}) and (\ref{eq7a}) the skewness parameters of the EoS along the $x$ and $y$ axes should be equal; i.e.,
\begin{equation}
\gamma=\delta.
\end{equation}
This shows that Bianchi type II spacetime (\ref{eq1}) does not allow the pressures of the DE fluid to be different along $x$ and $y$ directions as long as the shear stress of the overall EMT is null (i.e., as long as $T^{2}_{3}=0$, which gives $G^{2}_{3}=0$ (\ref{eq7a})). This is a result similar to the one discussed in a recent paper \cite{Saha11} by Saha. Hence, although the metric we consider is totally anisotropic, the anisotropy of the DE is constrained to be axially symmetric, as long as the overall EMT possesses zero shear stress.

\section{Model for power-law volumetric expansion}
\label{sec:3}

Using the equations (\ref{eq16}), (\ref{eq12}) and (\ref{eq19}), we obtain the directional scale factors as follows:
\begin{equation}
\label{eq20} A = (n \ell t + c_{1})^{\frac{3m}{n(m + 1)}} \;,
\end{equation}
\begin{equation}
\label{eq21}
B = c_{4}(n \ell t + c_{1})^{\frac{3m}{n(m + 1)}}\exp{\left[\frac{(m+1)^{2}}{2\ell ^{2}M}(n\ell t +c_{1})^{\frac{6(m-1)}{n(m+1)}+2 }+\frac{c_{3} }{ \ell (n-3)} (n\ell t + c_{1})^{\frac{n - 3}{n} }\right]} \;,
\end{equation}
\begin{equation}
\label{eq22}
C = c_{4}^{-1}(n \ell t + c_{1})^{\frac{3(1-m)}{n(m+ 1)}} \exp{\left[-\frac{(m+1)^{2}}{2\ell ^{2}M}(n\ell t +c_{1})^{\frac{6(m-1)}{n(m+1)}+2 }-\frac{c_{3} }{ \ell (n-3)} (n\ell t + c_{1})^{\frac{n - 3}{n} }\right]} \;,
\end{equation}
where $M=(9m-3+mn+n)(3m-3+mn+n)$ and $\;n\neq3$. In the special case $n=3$, the directional scale factors are as follows:
\begin{equation}
A = (3 \ell t + c_{1})^{\frac{m}{m + 1}} \;,
\end{equation}
\begin{equation}
B = c_{4}(3 \ell t + c_{1})^{\frac{3m\ell+c_{3}(m +1)}{3\ell(m + 1)}} \exp{\left[\frac{(m+1)^{2}}{144\ell ^{2}m^{2}}(3\ell t + c_{1})^{\frac{4m}{m+1}}\right]} \;,
\end{equation}
\begin{equation}
C = c_{4}^{-1}(3 \ell t
+c_{1})^{\frac{3(1-m)\ell-c_{3}(m + 1)}{3\ell(m +
1)}}\exp{\left[-\frac{(m+1)^{2}}{144\ell ^{2}m^{2}}(3\ell t
+c_{1})^{\frac{4m}{m+1} }\right]} \;.
\end{equation}
Using the directional scale factors, we obtain the directional Hubble parameters as follows:
\begin{equation}
\label{eq28}
H_{x}  = \frac{3m\ell}{m+1}(n\ell t+c_{1})^{-1},
\end{equation}
\begin{equation}
\label{eq29}
H_{y} = \frac{3m\ell}{m+1}(n\ell t+ c_{1})^{-1} +\frac{m+1}{\ell (9m-3+mn+n)}(n\ell t + c_{1})^{-\frac{6(1 - m)} {n(m + 1)}+1}+ c_{3} (n\ell t +c_{1})^{-\frac{3}{n}},
\end{equation}
\begin{equation}
\label{eq30}
H_{z} =\frac{3(1-m)\ell}{m+1}(n\ell t + c_{1})^{-1} -\frac{m+1}{\ell (9m-3+mn+n)}(n\ell t + c_{1})^{-\frac{6(1 - m)} {n(m + 1)}+1}- c_{3} (n\ell t +c_{1})^{-\frac{3}{n}}.
\end{equation}
The anisotropy parameter of the expansion is also obtained as follows:
\begin{eqnarray}
\Delta &=& \frac{2\sigma^{2}}{3H^{2}}=\frac{2(2m-1)^{2}}{(m+1)^{2}}
-\frac{15-33m-mn-n}{6\ell ^{2}(9m-3+mn+n)}(n\ell t +
c_{1})^{-\frac{6(1 - m)} {n(m + 1)}+2}  \nonumber\\
& &-\frac{2c_{3}(1-2m)}{\ell(m+1)}(n\ell t +c_{1})^{-\frac{3}{n}+1}+
\frac{2(m+1)^{2}}{3\ell^{4}(9m-3+mn+n)^{2}}(n\ell t +
c_{1})^{-\left[\frac{12(1 - m)} {n(m + 1)}-4\right]} \nonumber\\
& &+ \frac{4c_{3}(m+1)}{3\ell^{3}(9m-3+mn+n)}(n\ell t+
c_{1})^{-\left[\frac{3(3 - m)} {n(m + 1)}-3\right]}+
\frac{2c_{3}^{2}}{3\ell^{2}}(n\ell t + c_{1})^{-\frac{6}{n}+2}.
\end{eqnarray}

The energy density of the matter fluid is obtained as follows:
\begin{equation}
\label{21} \rho^{({\rm{m}})}=c_{0}(n\ell
t+c_{1})^{\frac{-3(1+w^{({\rm{m}})})}{n}}.
\end{equation}

The energy density, the EoS parameter on the $z$ axis and the skewness of the EoS parameter of the fluid we consider in the context of DE, respectively, are obtained as follows:
\begin{eqnarray}
\label{22}
\rho^{({\rm{de}})} &=& \frac{9\ell^{2}m(2-m)}{(m + 1)^{2}} (n \ell t +
c_{1})^{-2}+\frac{15-33m-mn-n}{4(9m-3+mn+n)}(n\ell t +
c_{1})^{-\frac{6(1 - m)} {n(m + 1)}}  \nonumber\\
& &+\frac{3c_{3}\ell(1-2m)}{m+1}(n\ell t +c_{1})^{-\frac{3}{n}-1}-
\frac{(m+1)^{2}}{\ell^{2}(9m-3+mn+n)^{2}}(n\ell t +
c_{1})^{-\left[\frac{12(1 - m)} {n(m + 1)}-2\right]} \nonumber\\
& &- \frac{2c_{3}(m+1)}{\ell(9m-3+mn+n)}(n\ell t+
c_{1})^{-\left[\frac{3(3 - m)} {n(m + 1)}-1\right]}- c_{3}^{2}(n\ell
t + c_{1})^{-\frac{6}{n}}-c_{0}(n\ell
t+c_{1})^{\frac{-3(1+w^{({\rm{m}})})}{n}},
\end{eqnarray}

\begin{eqnarray}
w^{({\rm{de}})} &=& \frac{1}{\rho^{({\rm{de}})}}\Bigg[\frac{3\ell^{2}[n(m+1)- 3(m^{2}-m+1)]}{(m + 1)^{2}}
(n \ell t + c_{1})^{-2}+\frac{3(m+1)(n+1)}{4(9m-3+mn+n)}(n\ell t +
c_{1})^{-\frac{6(1 - m)} {n(m + 1)}} \nonumber\\
& & +\frac{3c_{3}\ell(1-2m)}{m+1}(n\ell t +c_{1})^{-\frac{3}{n}-1}-
\frac{(m+1)^{2}}{\ell^{2}(9m-3+mn+n)^{2}}(n\ell t +
c_{1})^{-\left[\frac{12(1 - m)} {n(m + 1)}-2\right]}\nonumber\\
& &- \frac{2c_{3}(m+1)}{\ell(9m-3+mn+n)}(n\ell t +
c_{1})^{-\left[\frac{3(3 - m)} {n(m + 1)}-1\right]}- c_{3}^{2}(n\ell
t + c_{1})^{-\frac{6}{n}}-c_{0}w^{({\rm{m}})}(n\ell
t+c_{1})^{\frac{-3(1+w^{({\rm{m}})})}{n}}\Bigg],
\end{eqnarray}

\begin{equation}
\label{eq26} \delta =
\frac{1}{\rho^{({\rm{de}})}}\Bigg[\frac{3\ell^{2}(2m-1)(3 - n)}{m
+1}(n\ell t + c_{1})^{-2} + 2(n\ell t + c_{1})^{-\frac{6(1- m)}{n(m+
1)}}\Bigg] \;.
\end{equation}

The density parameter of the matter fluid  $\Omega^{({\rm{m}})}$  and
the density parameter of DE fluid $\Omega^{({\rm{de}})}$ may also be given
as follows:
\begin{equation}\label{26}
\Omega^{({\rm{m}})}=\frac{\rho^{({\rm{m}})}}{3H^{2}}=\frac{c_{0}}{3\ell^{2}}(n\ell
t+c_{1})^{\frac{2n-3(1+w^{({\rm{m}})})}{n}},
\end{equation}

\begin{eqnarray}\label{27}
\Omega^{({\rm{de}})} &=& \frac{\rho^{({\rm{de}})}}{3H^{2}}=\frac{3m(2-m)}{(m +
1)^{2}} +\frac{15-33m-mn-n}{12\ell^{2}(9m-3+mn+n)}(n\ell t +
c_{1})^{-\frac{6(1 - m)} {n(m + 1)}+2}-
\frac{c_{3}^{2}}{3\ell^{2}}(n\ell t +
c_{1})^{-\frac{6}{n}+2}  \nonumber\\
& &+\frac{c_{3}(1-2m)}{\ell(m+1)}(n\ell t +c_{1})^{-\frac{3}{n}+1}-
\frac{(m+1)^{2}}{3\ell^{4}(9m-3+mn+n)^{2}}(n\ell t +
c_{1})^{-\left[\frac{12(1 - m)} {n(m + 1)}-4\right]} \nonumber\\
& &- \frac{2c_{3}(m+1)}{3\ell^{3}(9m-3+mn+n)}(n\ell t+
c_{1})^{-\left[\frac{3(3 - m)} {n(m +
1)}-3\right]}-\frac{c_{0}}{3\ell^{2}}(n\ell
t+c_{1})^{\frac{-3(1+w^{({\rm{m}})})}{n}+2}.
\end{eqnarray}

Adding \eqref{26} and \eqref{27}, on defining total density parameter as the sum of the density parameters of the matter and DE fluids, we obtain the following:
\begin{equation}\label{29}
\Omega^{({\rm{T}})}\equiv\Omega^{({\rm{m}})}+\Omega^{({\rm{de}})} =\frac{3m(2-m)+(2m-1)^{2}}{(m +1)^{2}}-\frac{1}{2}\Delta.
\end{equation}

One may check that the solution satisfies the EMT conservation equations (\ref{2.12}) and (\ref{2.13}).

The initial time of the Universe is $t=-c_{1}/nl$, which can be shifted to $t=0$ by choosing $c_{1}=0$. Because this study is in the context of DE, we can conveniently consider only relatively later times of the Universe rather than early Universe in our discussion about the results.

We have already discussed that $m=\frac{1}{2}$ is the necessary but insufficient condition for approaching isotropy continuously and one may, here, check that  this model satisfies all the conditions for approaching isotropy continuously provided that $m=\frac{1}{2}$ and $n<1$ at the same time. In other words, the accelerating models for $m=\frac{1}{2}$ approaches isotropy continuously and also assures the positivity condition for the energy densities of the fluids in the late Universe. We may elaborate the late time behavior of the expansion anisotropy $\Delta$. One may check that the anisotropy of the expansion does not diverge as $t\rightarrow \infty$ provided that $n<1$ and $m<\frac{3-n}{3+n}$. In our proceeding discussion in this section, we consider only the cases that satisfy these two conditions at the same time, and for sufficiently later times of the Universe (for instance, today) we get the following results. Expansion anisotropy converges to a constant value for large $t$ values:
\begin{equation}
\Delta \approx \frac{2(2m-1)^{2}}{(m +1)^{2}}.
\end{equation}
One may easily observe that as would be expected from an accelerating model with $m=\frac{1}{2}$, $\Delta\approx 0$ for large $t$ values. Otherwise, $\Delta$ converges to a finite non-zero value for large $t$ values. For instance, Campanelli et al. \cite{Campanelli11a} gives $\mid\sqrt{\Delta}\mid\sim10^{-5}$ for the present Universe from WMAP data, which corresponds to $m\sim 0.5\pm 0.0000053$ in our model.

\begin{figure}[h]
\psfrag{A1}[b][b]{$\Delta$}
\psfrag{t}[b][b]{$t$}
\centering
\includegraphics[width=9cm]{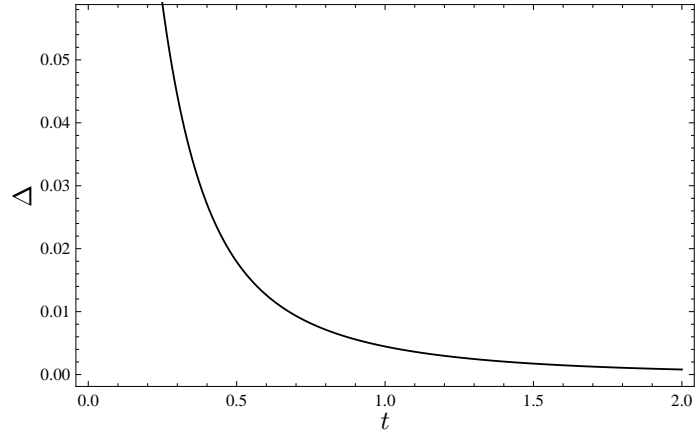}
\caption{The anisotropy of the expansion $\Delta$ versus the cosmic time $t$. }
\label{fig:anisotropy} 
\end{figure}

\begin{figure}[h]
\psfrag{Density Parameters}[b][b]{Density Parameters}
\psfrag{t}[b][b]{$t$}
\centering
\includegraphics[width=9cm]{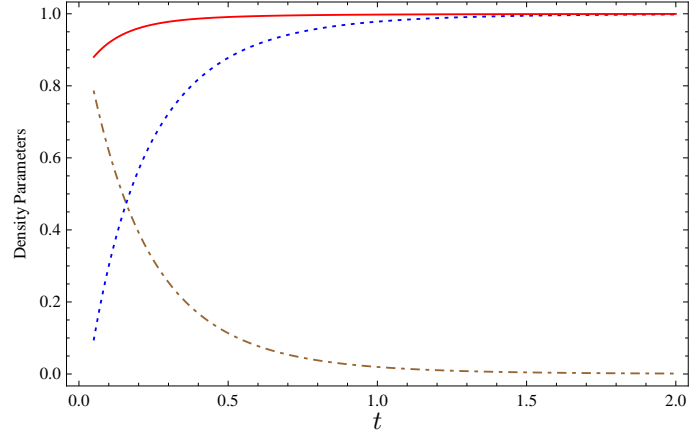}
\caption{The density parameters of the dark energy fluid $\Omega^{({\rm{de}})}$ (dotted curve) and matter fluid $\Omega^{({\rm{m}})}$ (dot-dashed curve) and the total density parameter $\Omega^{({\rm{T}})}$ (solid curve) versus the cosmic time $t$. }
\label{fig:dp} 
\end{figure}

The behaviors of the energy densities of the fluids also need attention to decide whether our model is realistic or not.  Obviously for large $t$ values, the density parameter of the matter fluid is
\begin{equation}
\Omega^{({\rm{m}})}\approx 0
\end{equation}
independent of whether the Universe exhibits accelerating volumetric expansion or not. One may observe that, on the other hand, DE dominates the matter fluid for large $t$ values for accelerating models and its density parameter approximates to the following finite value:
\begin{equation}
\Omega^{({\rm{de}})} \approx \frac{3m(2-m)}{(m +1)^{2}}.
\end{equation}
Hence, the total density parameter also converges to
\begin{equation}
\Omega^{({\rm{T}})} \approx \frac{3m(2-m)}{(m +1)^{2}}.
\end{equation}
Hence, for sufficiently later times of the Universe, DE energy fluid dominates the matter fluid provided the above mentioned conditions, and in particular case for $m=\frac{1}{2}$, we have $\Omega^{({\rm{T}})}\approx1$. Thus, for $m=\frac{1}{2}$ and $n<1$, our model predicts a spatially flat and isotropic Universe at late times. For a demonstration, we depict the anisotropy of the expansion in Fig. \ref{fig:anisotropy} and density parameters in Fig. \ref{fig:dp}. In these figures, also in all other figures, we consider $m=\frac{1}{2}$,  $n=0.27$ ($q=-0.73$) that corresponds to the value obtained from observations for the present Universe (e.g., see \cite{Cunha,Li2011}), $w^{({\rm{m}})}=0$ since the most dominant ingredient after the DE in the present Universe is the pressureless matter. On the other hand, for the other constants we have selected the values as $\ell=2$, $c_{0}=12$, $c_{1}=1$, $c_{3}=1$, which give the true beginning of the universe $t=-1.85$\footnote{Here and in the figures, for convenience, we have not used dimensions for the cosmic time $t$, since one can always set the scale and the dimension of the cosmic time so as it to correspond to the true cosmic time.}. However, because we are not interested in the early times of the Universe but the later times, for convenience, we plot the graphs from $t=0$ to $t=2$ that can show us the generic behaviors of the parameters for $m=\frac{1}{2}$ and $q=-0.73$. From Fig. \ref{fig:anisotropy}, one may observe that the anisotropy parameter $\Delta$ \eqref{eq26} has larger values at earlier times of the Universe and it decreases monotonically to zero and flattens as $t$ increases ($\Delta\rightarrow 0$ and $\dot{\Delta}\rightarrow 0$ as $t\rightarrow \infty$). One may also check that $\Delta$ diverges at the true beginning of the Universe. Moreover, it dominates the physical sources in the sufficiently early times of the Universe. Therefore, in view of equation \eqref{29}, the total density parameter (i.e., density parameter of the matter field plus the density parameter of the DE) curve goes down as we go back into time (see Fig. \ref{fig:dp}).

\begin{figure}[h]
\psfrag{kin}[b][b]{\footnotesize{Energy Densities}}
\psfrag{t}[b][b]{$t$}
\centering
\includegraphics[width=9cm]{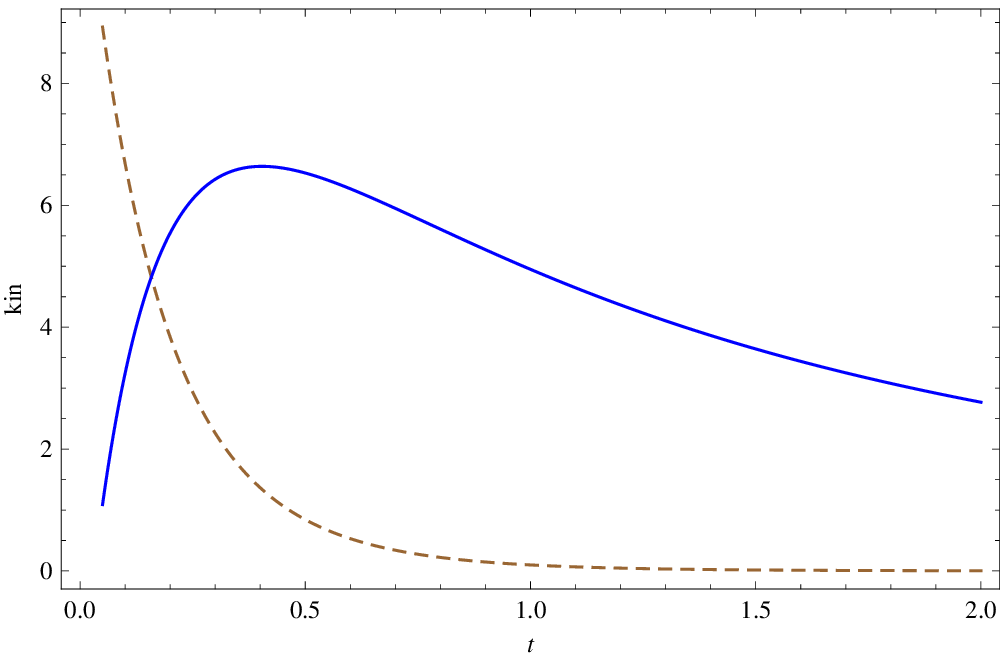}
\caption{The energy densities of the matter fluid $\rho^{({\rm{m}})}$ (dashed curve) and dark energy fluid $\rho^{({\rm{de}})}$ (solid curve) versus the cosmic time $t$. }
\label{fig:rho}
\end{figure}

\begin{figure}[h]
\psfrag{EoS parameters}[b][b]{EoS parameters}
\psfrag{t}[b][b]{$t$}
\centering
\includegraphics[width=9cm]{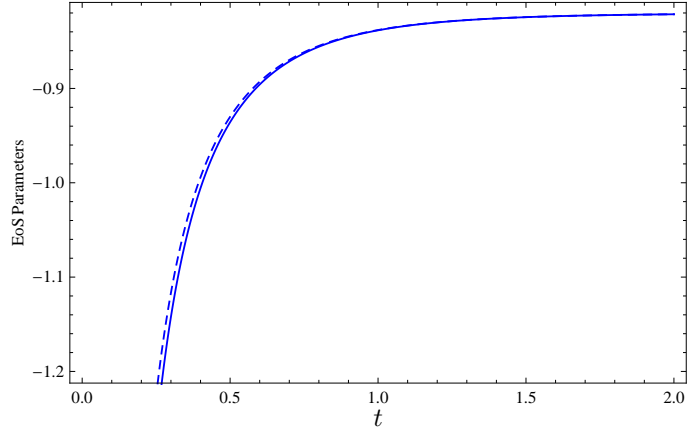}
\caption{The directional equation of state (EoS) parameters of the dark energy fluid along the $x$ and $y$ axes $w^{({\rm{de}})}_{x}=w^{({\rm{de}})}_{y}$ (dashed cruve) and along the $z$ axis $w^{({\rm{de}})}=w^{({\rm{de}})}_{z}$ (solid cruve) versus the cosmic time $t$.}
\label{fig:eoss} 
\end{figure}

\begin{figure}[h]
\psfrag{Deltaw}[b][b]{$\delta/w^{({\rm{de}})}$}
\psfrag{t}[b][b]{$t$}
\centering
\includegraphics[width=9cm]{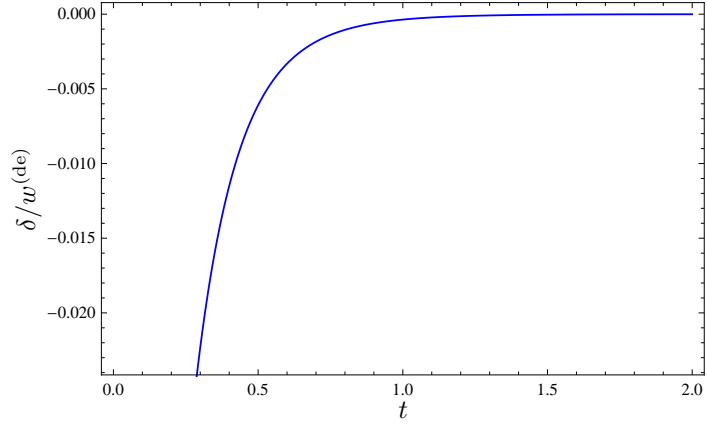}
\caption{Anisotropy of the dark energy fluid $\delta/w^{({\rm{de}})}$ versus the cosmic time $t$}
\label{fig:ande}
\end{figure}

We would also like to discuss on the dynamics of the fluid we consider in the context of DE little bit more under the conditions $n<1$ and $m<\frac{3-n}{3+n}$. The energy density of the matter fluid decreases monotonically as the Universe evolves, while the energy density of the DE fluid increases until a certain time of the Universe and eventually converges to a finite value depending on the value of $m$ in the later Universe (see Fig. \ref{fig:rho}). The directional EoS parameters of the DE eventually evolve into the quintessence region. One may observe that both $w^{({\rm{de}})}_{x}=w^{({\rm{de}})}_{y}$ and $w^{({\rm{de}})}=w^{({\rm{de}})}_{z}$ start in the phantom region and evolve into the quintessence region (see Fig. \ref{fig:eoss}). This is consistent with the observations, for instance, with the range $(-1.45,-0.74)$ given by Melchiorri et al. \cite{Melchiorri}. The anisotropy of the DE can be measured by $\frac{\delta}{w^{({\rm{de}})}}$ and it asymptotically tends to $\frac{(2m-1)(3 - n)(m + 1)}{n(m+1)-3(m^2 -m+1)}$ for large $t$ values. In particular, $\frac{\delta}{w^{({\rm{de}})}}$ becomes zero at late times for the accelerating models with $m=\frac{1}{2}$ (see Fig. \ref{fig:ande}), which in turn implies that the anisotropy of DE goes off during the cosmic evolution.

\vspace{2cm}

\vspace{2cm}

\pagebreak

\section{Model for exponential volumetric expansion}
\label{sec:4}
Solving the equations (\ref{eq17}), (\ref{eq12}) and (\ref{eq19}), we obtain the directional scale factors as follows:
\begin{equation}
\label{eq39} A = c_{2}^{\frac{3m}{m+1}}\exp{\left(\frac{3m\ell}{m+1}t\right)},
\end{equation}
\begin{equation}
\label{eq40} B = c_{4}c_{2}^{\frac{3m}{m+1}}\exp{\left[\frac{3m\ell}{m+1}t +\frac{c_{2}^{\frac{6(m-1)}{m+1}}(m+1)^{2}}{18\ell^{2}(3m-1)(m-1)}e^{\frac{6\ell(m-1)}{m+1}t}- \frac{c_{3}} {3\ell c_{2}^{3}}e^{-3\ell t}\right]},
\end{equation}
\begin{equation}
\label{eq41} C = c_{4}^{-1}c_{2}^{\frac{3(1-m)}{m+1}}\exp{\left[\frac{3(1-m)\ell}{m+1}t-\frac{c_{2}^{\frac{6(m-1)}{m+1}}(m+1)^{2}}{18\ell^{2}(3m-1)(m-1)}e^{\frac{6\ell(m-1)}{m+1}t}+\frac{c_{3}} {3\ell c_{2}^{3}}e^{-3\ell t}\right]}.
\end{equation}

The directional Hubble parameters are obtained as follows:
\begin{equation}
\label{eq51} H_{x} =  \frac{3m\ell}{m + 1},
\end{equation}
\begin{equation}
\label{eq52} H_{y} =  \frac{3m\ell}{m+1}
+\frac{c_{2}^{\frac{6(m-1)}{m+1}}(m+1)}{3\ell(3m-1)}e^{\frac{6\ell(m-1)}{m+1}t}+
\frac{c_{3}} {c_{2}^{3}}e^{-3\ell t},
\end{equation}
\begin{equation}
\label{eq53} H_{z} =  \frac{3\ell(1-m)}{m+1}
-\frac{c_{2}^{\frac{6(m-1)}{m+1}}(m+1)}{3\ell(3m-1)}e^{\frac{6\ell(m-1)}{m+1}t}-
\frac{c_{3}} {c_{2}^{3}}e^{-3\ell t}.
\end{equation}

The expansion anisotropy parameter is obtained as follows:
\begin{eqnarray}\label{eq48}
\Delta &=& \frac{2(2m-1)^{2}}{(m + 1)^{2}}
-\frac{c_{2}^{\frac{6(m-1)}{m+1}}(5-11m)}{6\ell^{2}(3m-1)}e^{\frac{6\ell(m-1)}{m+1}t}
+\frac{4c_{3}c_{2}^{\frac{3(m-3)}{m+1}}(m+1)}{9\ell^{3}(3m-1)}e^{\frac{3\ell(m-3)}{m+1}t}
\nonumber\\
& &-\frac{2c_{3}(1-2m)} {\ell^{2}c_{2}^{3}(m+1)}e^{-3\ell t}
+\frac{2c_{2}^{\frac{12(m-1)}{m+1}}(m+1)^{2}}{27\ell^{4}(3m-1)^{2}}e^{\frac{12\ell(m-1)}{m+1}t}
+\frac{2c_{3}^{2}}{3\ell^{2}c_{2}^{6}}e^{-6\ell t}.
\end{eqnarray}

The energy density of the matter fluid reads as
\begin{equation}
\rho^{({\rm{m}})}=c_{0}c_{2}^{-3(1+w^{({\rm{m}})})}e^{-3\ell(1+w^{({\rm{m}})})t},
\end{equation}

The energy density, the EoS parameter on the $z$ axis and the skewness of the EoS parameter of the fluid we consider in the context of DE, respectively, are obtained as follows:
\begin{eqnarray}\label{eq44}
\rho^{({\rm{de}})} &=& \frac{9\ell^{2}m(2-m)}{(m + 1)^{2}}
+\frac{c_{2}^{\frac{6(m-1)}{m+1}}(5-11m)}{4(3m-1)}e^{\frac{6\ell(m-1)}{m+1}t}
-\frac{2c_{3}c_{2}^{\frac{3(m-3)}{m+1}}(m+1)}{3\ell(3m-1)}e^{\frac{3\ell(m-3)}{m+1}t}+\frac{3c_{3}(1-2m)}
{c_{2}^{3}(m+1)}e^{-3\ell t}
\nonumber\\
& &
-\frac{c_{2}^{\frac{12(m-1)}{m+1}}(m+1)^{2}}{9\ell^{2}(3m-1)^{2}}e^{\frac{12\ell(m-1)}{m+1}t}
-\frac{c_{3}^{2}}{c_{2}^{6}}e^{-6\ell
t}-c_{0}c_{2}^{-3(1+w^{({\rm{m}})})}e^{-3\ell(1+w^{({\rm{m}})})t},
\end{eqnarray}

\begin{eqnarray}
\label{eq43}
w^{({\rm{de}})} &=&
\frac{1}{\rho^{({\rm{de}})}}\Bigg[-\frac{9\ell^{2}(m^{2}-m+1)}{(m + 1)^{2}}
+\frac{c_{2}^{\frac{6(m-1)}{m+1}}(m+1)}{4(3m-1)}e^{\frac{6\ell(m-1)}{m+1}t}
-\frac{2c_{3}c_{2}^{\frac{3(m-3)}{m+1}}(m+1)}{3\ell(3m-1)}e^{\frac{3\ell(m-3)}{m+1}t}
\nonumber\\
& & +\frac{3c_{3}(1-2m)} {c_{2}^{3}(m+1)}e^{-3\ell
t}-\frac{c_{2}^{\frac{12(m-1)}{m+1}}(m+1)^{2}}{9\ell^{2}(3m-1)^{2}}e^{\frac{12\ell(m-1)}{m+1}t}
-\frac{c_{3}^{2}}{c_{2}^{6}}e^{-6\ell
t}-c_{0}c_{2}^{-3(1+w^{({\rm{m}})})}e^{-3\ell(1+w^{({\rm{m}})})t}\Bigg],
\end{eqnarray}

\begin{equation}
\label{eq45}
\delta = \frac{1}{\rho^{({\rm{de}})}}\Bigg[\frac{9\ell^{2}(2m-1)}{m +1}+2c_{2}^{\frac{6(m - 1)}{m + 1}}e^{\frac{6\ell(m-1)}{m+1}t}\Bigg].
\end{equation}

The density parameters of the matter fluid and the DE fluid are obtained as follows:
\begin{equation}\label{18}
\Omega^{({\rm{m}})}=\frac{c_{0}c_{2}^{-3(1+w^{({\rm{m}})})}}{3\ell^{2}}e^{-3\ell(1+w^{({\rm{m}})})t},
\end{equation}

\begin{eqnarray}
\Omega^{({\rm{de}})} &=& \frac{3m(2-m)}{(m + 1)^{2}}
+\frac{c_{2}^{\frac{6(m-1)}{m+1}}(5-11m)}{12\ell^{2}(3m-1)}e^{\frac{6\ell(m-1)}{m+1}t}
-\frac{2c_{3}c_{2}^{\frac{3(m-3)}{m+1}}(m+1)}{9\ell^{3}(3m-1)}e^{\frac{3\ell(m-3)}{m+1}t}+\frac{c_{3}(1-2m)}
{\ell^{2}c_{2}^{3}(m+1)}e^{-3\ell t}
\nonumber\\
& &
-\frac{c_{2}^{\frac{12(m-1)}{m+1}}(m+1)^{2}}{27\ell^{4}(3m-1)^{2}}e^{\frac{12\ell(m-1)}{m+1}t}
-\frac{c_{3}^{2}}{3\ell^{2}c_{2}^{6}}e^{-6\ell
t}-\frac{c_{0}c_{2}^{-3(1+w^{({\rm{m}})})}}{3\ell^{2}}e^{-3\ell(1+w^{({\rm{m}})})t}.
\end{eqnarray}

The total density parameter is obtained as follows:
\begin{equation}
\Omega^{({\rm{T}})}=\frac{3m(2-m)+(2m-1)^{2}}{(m +1)^{2}}-\frac{1}{2}\Delta.
\end{equation}

One may check that the solution satisfies the EMT conservation equations (\ref{2.12}) and (\ref{2.13}), as expected.

In this model for $n=0$, the DP of the volumetric expansion is $q=-1$, which leads to $dH/dt=0$ that implies the greatest value of Hubble's parameter and the fastest rate of expansion for a Universe expanding forever. This is one of the possible futures of our Universe. Therefore, the derived model can be utilized to describe the dynamics of the future of the present Universe. So, in what follows, we emphasize upon the late time behavior of the derived model. One may check that only in the models for $m<1$, expansion anisotropy $\Delta$ does not diverge as $t\rightarrow \infty$ but converges to a constant, and the energy density of the fluid we consider in the context of DE does not evolve to negative values.

Hence, the detailed picture of sufficiently older Universe is as follows; for $m<1$ we obtain
\begin{eqnarray}
\Delta \approx \frac{2(2m-1)^{2}}{(m + 1)^{2}},
\end{eqnarray}
\begin{eqnarray}
 \rho^{({\rm{m}})} \approx 0,
 \end{eqnarray}
\begin{eqnarray}
 \rho^{({\rm{de}})} \approx \frac{9\ell^{2}m(2-m)}{(m + 1)^{2}}, \qquad w^{({\rm{de}})} \approx -\frac{m^{2}-m+1}{m(2-m)}, \qquad\delta \approx \frac{(2m-1)(m+1)}{m(2-m)}, \qquad  \frac{\delta}{w^{({\rm{de}})}} \approx \frac{(2m-1)(m+1)}{m^2 -m+1}
 \end{eqnarray}
 \begin{eqnarray}
\Omega^{({\rm{T}})} \approx\frac{3m(2-m)}{(m + 1)^{2}},
\end{eqnarray}
and for the particular case $m=\frac{1}{2}$, we further obtain
\begin{eqnarray}
\Delta \approx0,
\end{eqnarray}
\begin{eqnarray}
\rho^{({\rm{m}})}\approx 0,
 \end{eqnarray}
\begin{eqnarray}
\rho^{({\rm{de}})}\approx 3\ell^{2}, \qquad w^{({\rm{de}})}\approx-1, \qquad \delta \approx 0,\qquad \frac{\delta}{w^{({\rm{de}})}} \approx 0
 \end{eqnarray}
 \begin{eqnarray}
\Omega^{({\rm{T}})}\approx 1.
\end{eqnarray}
One may observe that if $m=\frac{1}{2}$, as $t\rightarrow \infty$, the expansion anisotropy converges to zero, the energy density of the matter fluid converges to zero, the fluid we consider in the context of DE mimics the well known positive cosmological constant (i.e., DE fluid isotropizes and its energy density takes a positive finite value), the total density parameter converges to unity, which means that the Universe becomes spatially isotropic and flat. Thus, the model for $m=\frac{1}{2}$ is well behaved and satisfies all the conditions for approaching isotropy continuously. In short, for $m=\frac{1}{2}$ our model that exhibits exponential volumetric expansion can evolve to a phase where it is observationally indistinguishable from the well known solution of the Einstein field equations in the framework of RW spacetime in the presence of positive cosmological constant, which is also known as de Sitter Universe and is the future of the Universe in the standard $\Lambda$CDM model.

\section{Concluding Remarks}
\label{conclusion}

We obtain cosmological models that can exhibit rich variety of behaviors in the presence of two minimally interacting fluids (a perfect fluid as the matter fluid and a hypothetical anisotropic fluid in the context of DE) within the framework of spatially homogenous but totally anisotropic and non-flat Bianchi type II spacetime in GR constrained with two kinematical ans\"{a}tze. As the first kinematical ansatz, we assumed the constant mean DP whose negative values give accelerating volumetric expansion. Although it is incapable of representing the whole or a long period of history of the Universe at once, it can still be considered as an approximation for representing a particular period of the Universe, for instance, the vicinity of the present time and the very late times of the Universe. As the second kinematical ansatz, we assumed one of the component of the shear tensor is proportional to the mean Hubble parameter.

We showed that although the Bianchi II spacetime we considered is totally anisotropic, it admits only an axially symmetric anisotropic pressure of a comoving fluid as long as shear tension does not exist, independently of the ans\"{a}tze we considered. We also showed that, for the models that exhibit accelerating volumetric expansion ($n<1$), there are well-behaved cases where the anisotropy of the expansion converges to finite values (including zero) and the fluid we consider in the context of DE yields positive values for its energy density in the late Universe.

In the particular case with $m=\frac{1}{2}$ among the well-behaved cases, the anisotropy of the expansion converges to zero continuously, the total density parameter converges to unity and the anisotropy of the DE converges to zero, which means that our model becomes indistinguishable from a spatially anisotropic and flat Universe in the presence of isotropic fluids. Further, in the case for $n=0$ and $m=1/2$, the fluid we consider in the context of DE mimics the cosmological constant and the Universe evolves to the well known de Sitter Universe.

Yielding anisotropic pressure, the fluid we consider in the context of DE, can produce results that can be produced in the presence of isotropic fluid in accordance with the $\mathrm{\Lambda}$CDM cosmology. However, our model gives additional opportunities by being able to allow kinematics that cannot be produced in the presence of fluids that yield only isotropic pressure. There are also well behaved cases where the anisotropy of the expansion and the anisotropy of the fluid do not cease but converge to non-zero finite values in the late Universe. Hence, the dominance of such anisotropic DE candidates can distort the isotropy of the Universe, which is believed to be gained in the inflationary era in the early Universe, with required limits and characteristics. For instance, recently, Campanelli et al. \cite{Campanelli11a} investigated the possibility that an anisotropic DE component could generate an ellipsoidal Universe with the correct characteristics to explain the low quadrupole in the CMB fluctuations and proposed that an ellipsoidal Universe allowed better matches with the large-scale CMB anisotropy data. They give $\mid\sqrt{\Delta}\mid\sim10^{-5}$ for the anisotropy of the expansion of the present Universe from WMAP data, which corresponds to $m\sim 0.5\pm 0.0000053$ in our model.

Our models where the Universe accelerates with an expansion anisotropy that can evolve to a finite value (including zero) at late times would be of interest to many readers since it can generate isotropic expansion as would be expected in standard cosmology as well as the non-divergent anisotropic expansion in accordance with the discussions on whether isotropic expansion is good enough to explain cosmological observations or not.

\begin{center}
\textit{Acknowledgments}
\end{center}
\"{O}.A. acknowledges the support he is receiving from the Turkish Academy of Sciences (T\"{U}BA) and Ko\c{c} University. S.K. acknowledges the encouragement for research activities by the authorities of Delhi Technological University.

\end{document}